\documentclass[12pt]{article}
\pdfoutput=1
\usepackage[a4paper]{geometry}
 
\usepackage[T1]{fontenc} 
\usepackage{comment}

\usepackage{jheppub,hyperref,float,array,adjustbox,mathtools,physics, xcolor}

\usepackage{xcolor,tikz,pgfplots,amsmath,amsfonts}
\usepackage{graphicx}
\usetikzlibrary{matrix,calc,positioning,decorations.markings,decorations.pathmorphing,decorations.pathreplacing}%tikzmark,
\usetikzlibrary{arrows,cd}
\usetikzlibrary{shapes.misc}
\usetikzlibrary {shapes.geometric}
\usetikzlibrary {shapes.multipart}
\usetikzlibrary{decorations.markings}
\usetikzlibrary{backgrounds}
\usepackage{bbding}

\usepackage{tikz}
\usetikzlibrary{automata,positioning,calc}
\usetikzlibrary{decorations.markings}
\tikzset{->-/.style={decoration={markings, mark=at position #1 with {\arrow{>}}},postaction={decorate}}}
\tikzset{-<-/.style={decoration={markings, mark=at position #1 with {\arrow{<}}},postaction={decorate}}}
\tikzset{auto shift/.style={auto=right,->, to path={ let \p1=(\tikztostart),\p2=(\tikztotarget), \n1={atan2(\y2-\y1,\x2-\x1)},\n2={\n1+180} in ($(\tikztostart.{\n1})!1mm!270:(\tikztotarget.{\n2})$) -- ($(\tikztotarget.{\n2})!1mm!90:(\tikztostart.{\n1})$) \tikztonodes}}}

\usetikzlibrary {shapes.geometric} 

\newcommand{\Pf}[1]{\,\mathrm{Pf} #1}
\newcommand{\USp}[1]{\mathrm{USp}(#1)}
\newcommand{\SU}[1]{\mathrm{SU}(#1)}
\newcommand{\U}[1]{\mathrm{U}(#1)}

\usepackage{float}

\title{
\begin{center}
Sporadic dualities from  tensor deconfinement
\end{center}
}

\author[a]{Antonio Amariti,}	
\author[b]{Fabio Mantegazza,}
\author[a,b]{Davide Morgante,}

\affiliation[a]{INFN, Sezione di Milano, Via Celoria 16, I-20133 Milano, Italy}
\affiliation[b]{Dipartimento di Fisica, Università degli studi di Milano, Via Celoria 16, I-20133, Milano, Italy}

\emailAdd{antonio.amariti@mi.infn.it}
\emailAdd{fabio.mantegazza2@studenti.unimi.it}
\emailAdd{davide.morgante@mi.infn.it}

\abstract{
In this paper we give a field theory explanation of  two confining dualities that have been proposed in the literature based on exact  results from supersymmetric localization.
The first confining model under investigation is 4d $\SU{N_c+1}$ SQCD with a conjugate rank-$2$ anti-symmetric tensor,  $N_c+3$ anti-fundamentals, $2N_c$ fundamentals and a superpotential that couples the anti-symmetric tensor and the fundamentals. 
The second confining model studied here is  $3d$ $\mathcal{N}=2$ $\USp{4}$ gauge SQCD with two fundamentals, two rank-$2$ anti-symmetric tensors and vanishing superpotential. 
Here we prove that these models are confining by  using the technique of deconfining the anti-symmetric tensors and then by flowing to the IR description by sequential dualities. 
As a bonus the analysis provides (alternative) proofs of the identities obtained from supersymmetric localization.
}

\begin{document}
\maketitle
\flushbottom
\allowdisplaybreaks 
\section{Introduction}

The low energy dynamics of UV free strongly coupled supersymmetric gauge theories 
can be often be simplified by the existence of infrared dualities. The dual descriptions
are in general associated to (more) weakly coupled QFTs, described by a different set of 
fields and interactions that share in the IR the same correlation functions for the physically observable conserved currents of the original description.
The prototypical example of such dualities is the electromagnetic duality and for this reason
the two dual models are usually referred to as the electric and the magnetic phase.

Restricting to cases with four supercharges the basic example of these dualities was discovered by Seiberg in \cite{Seiberg:1994pq} for $\SU{N_c}$ 4d SQCD with $N_f >N_c+1$ flavors and vanishing superpotential.
This duality has also a limiting case, where the magnetic description does not correspond to any gauge theory but to a WZ model consisting in a collection of mesons and baryons of the electric description, in addition to a (classical) constraint among them. In this case, corresponding to the choice $N_f = N_c+1$, the electric gauge theory confines without breaking the chiral symmetry (i.e. s-confines \cite{Csaki:1996sm}), and the magnetic theory describes the dynamics of the confined degrees of freedom, with a superpotential imposing the classical constraint on the moduli space.
There is also another confining case, corresponding to $\SU{N_c}$ 4d SQCD with $N_f=N_c$ flavors, where the low energy dynamics described by the mesons and the baryons requires a quantum constraint on the moduli space. Such constraint breaks the chiral symmetry and for this reason this case is referred to as confinement with chiral symmetry breaking.

This idea of confinement as a limiting case of a supersymmetric duality was then extended to various generalizations of Seiberg duality. Furthermore a full classification of s-confining gauge theories with vanishing superpotential was worked out in \cite{Csaki:1996zb} for theories with a single gauge group.
In this classification there are many models that do not correspond to any  limiting case of any known duality. Such models are characterized usually by the presence of matter fields in a rank-two tensor representation of the gauge group.
Despite the fact that gauge theories of this type do not have in general a Seiberg-like dual description, it has been shown in \cite{Bajeot:2022kwt} that the s-confining dualities can be derived using only Seiberg-(like) dualities thanks to the rank-$2$ tensor deconfining technique originally proposed in \cite{Berkooz:1995km} and subsequently generalized in \cite{Luty:1996cg}.\\
The technique consists of substituting a rank-$2$ tensor matter field with a bifundamental field charged also under another (auxiliary) confining gauge group, such to recover the original description once this new gauge group confines. After deconfining the rank-$2$ tensors it has been possible to apply sequences of Seiberg dualities (see \cite{Bottini:2022vpy} for a general construction) and than to recover the confined phase proposed in \cite{Csaki:1996zb}, using only the s-confining dualities  
of $\SU{N_c}$ with $N_c+1$ flavors of  \cite{Seiberg:1994pq} and $\USp{2N_c}$ SQCD with $2N_c+4$ fundamentals of \cite{Intriligator:1995ne}. 
This construction may require  further refinements for models with a superpotential deformation, due to the possible presence of an Higgsing that breaks partially or completely the gauge group (see \cite{Comi:2022aqo} for a general discussion).
\\ 
Recently new confining gauge theories have been obtained for 4d models with rank-$2$ tensors and non-vanishing superpotential \cite{Bajeot:2023gyl}.
Furthermore the deconfinement techniques have been applied to 3d $\mathcal{N}=2$ gauge theories \cite{Benvenuti:2020gvy}, 
where the zoo of confining gauge theories is richer, because of the presence of a dual photon and of a Coulomb branch. New confining dualities in this direction have been obtained in \cite{Benvenuti:2021nwt,Amariti:2022wae}.

In this paper we apply these techniques to a 4d and a 3d model that have been claimed to be confining because of integral identities in supersymmetric localization. We find a physical origin of these integral identities that allowed to state the new confining dualities, finding a field theoretical explanation for them.
The 4d duality under inspection corresponds to $\SU{N_c+1}$ SQCD, with a rank-$2$ conjugate anti-symmetric tensor, $N_c+3$ anti-fundamentals and $2N_c$ fundamentals. This theory is claimed to be confining if a cubic superpotential  between the anti-symmetric and the fundamentals is turned on.
Such claim was originally proposed in \cite{Spiridonov:2009za} based on the fact that the supersymmetric index\footnote{We refer to the supersymmetric index instead of the superconformal index because we focus on models that are out of the conformal windows. In the conformal window the two quantities coincide.} on $S^3 \times S^1$ \cite{Kinney:2005ej,Romelsberger:2005eg}  of this theory was computed exactly in \cite{spiridonov2004inversions}.
The final result has a field theory interpretation representing the low energy description of the baryons and mesons of the $\SU{N_c+1}$ gauge theory with the expected constraints from the truncation of the chiral ring and the moduli space.
This duality has been referred to as Spiridonov-Warnaar-Vartanov (SWV) duality in \cite{Nazzal:2021tiu}, where it  has been used in the study of 4d compactification of the 6d minimal (D,D) conformal matter theories on a punctured Riemann surface (see also \cite{Nazzal:2023bzu}).

Here we provide a physical derivation of the duality from the field theoretical perspective, by deconfining the rank-$2$ anti-symmetric tensor and  sequentially dualizing the gauge groups.
In the process we find that one of the steps requires a partial Higgsing, analogously  to the analysis recently  performed in \cite{Bajeot:2023gyl} for similar 4d confining dualities. The partial Higgsing is triggered in our case by an $\USp{2N_c}$   gauge group with $2N_c+2$ fundamentals, that confines breaking the chiral symmetry.
Furthermore, following the various steps on the supersymmetric index we provide an alternative derivation of the identity of \cite{spiridonov2004inversions}.

In the second part of the paper we study a 3d confining duality recently obtained in \cite{Okazaki:2023hiv}, corresponding to $\USp{4}$ with two rank-$2$ anti-symmetric tensors and two fundamentals. 
The existence of such a duality has been claimed by  extending to the 3d bulk a boundary duality constructed from $\mathcal{N}=(0,2)$ half-BPS boundary conditions in 3d $\mathcal{N}=2$. Again we deconfine the two rank-$2$ anti-symmetric tensors and then provide the sequential dualities leading to the final description in terms of the gauge singlets of the original model.

\section{The Spiridonov-Warnaar-Vartanov 4d duality}

In this section we derive the SWV duality   from a physical 
approach, by deconfining a rank-$2$  conjugate anti-symmetric tensor with an auxiliary symplectic gauge group and then by sequentially applying  IR dualities.
Actually referring to the last step in such a sequence as a duality is improper, because, as we will see in the following,  it corresponds to the case of a symplectic gauge theory that confines with a quantum constraint on the moduli space. The crucial aspect of this constraint is that it forces a Higgs mechanism on the leftover unitary gauge group, breaking it to a symplectic one, and assigning a superpotential mass term to some of the fields in the spectrum.
This leads to the final step of the construction, where one is left with an s-confining gauge theory (namely $\USp{2M}$ with $2M+4$ fundamentals). After confining this theory we eventually  find the expected WZ model, describing the magnetic phase of the SWV duality.

The analysis is supported at each step by the relative (integral) identities matching the 4d supersymmetric index. On one hand this corroborates the validity of the results 
and on the other hand it provides an alternative derivation of the integral identity discovered in \cite{spiridonov2004inversions}.

Let us start the analysis discussing the gauge theory that can be read from the Spiridonov-Warnaar identity \cite{spiridonov2004inversions}.
It consists in $\SU{N_c+1}$ SQCD with $N_c+3$ anti-fundamentals $Q_1$ and $2N_c$ fundamentals $Q_{2}$.
In addition there is a rank-$2$ anti-symmetric conjugate tensor $A$. 
This is a non-anomalous asymptotically free theory and it  becomes confining 
if the  superpotential deformation\footnote{In the rest of the paper the explicit contractions will be mostly understood.}
\begin{equation}
W_{ele}  = {Q_1}_i^{\alpha} J_{2N_c}^{ij} {Q_1}_j^\beta A_{\overline \alpha, \overline \beta}
\end{equation}
 is turned on. 
 This deformation is relevant for any value of $N_c$ \cite{Nazzal:2021tiu} and it breaks the $\SU{2N_c}$ flavor symmetry group into $\USp{2N_c}$.
 The representations of the fields and their charges under the gauge and the flavor groups are summarized in the following 
\begin{equation}
\label{repsSW1}
\begin{array}{c|c||cccc}
&\SU{N_c+1}&\USp{2N_c} & \SU{N_c+3}& \U{1} & \U{1}_R \\
\hline
Q_1 &\overline{T}_f & 1&T_f&1&0 \\
Q_2&T_f& T_f&1&-\frac{N_c+3}{2}&1 \\
A&\overline T_A&1&1&N_c+3&0\\
\end{array}
\end{equation}
The $S^3\times S^1$ supersymmetric index  of this model has been explicitly computed \cite{spiridonov2004inversions}.
The identity is $I_E = I_M$ with
\begin{equation}
\begin{split}
    I_E &= \frac{(p;p)^{N_c}_\infty (q;q)^{N_c}_\infty } {(N_c+1)!}
    \int_{\mathbb{T}^{N_c}} \prod_{1 \leq i<j \leq N_c+1} \frac{\Gamma(S z_i^{-1}  z_j^{-1}) }{\Gamma((z_i/z_j)^{\pm 1})}\\
    &\times\prod_{j=1}^{N_c+1} \prod_{k=1}^{N_c} \Gamma( t_k z_j) \cdot\frac{\prod_{m=1}^{N_c+3} \Gamma(s_m z_j^{-1})}{\prod_{k=1}^{N_c} \Gamma(S t_k z_j^{-1})}\frac{\dd{z_j}}{2 \pi i z_j}
\end{split}
\end{equation}
and
\begin{equation}
    I_M = \prod_{m=1}^{N_c+3} \prod_{k=1}^{N_c} \frac{\Gamma(s_m t_k)}{\Gamma(S s_m^{-1} t_k)} \prod_{1\leq l < m \leq N_c+3} \Gamma(S s_l^{-1} s_m^{-1} )
\end{equation}
with the constraint $S = \prod_{m=1}^{N_c+3} s_m$ imposed on the charges.

 It is possible to read from this identity that there are two contributions  
arising from the meson $M = Q_1 Q_2$ and the baryon $B = Q_1^{N_c+1}$.
The field content allows a non-vanishing superpotential of the form \cite{Nazzal:2021tiu}
\begin{equation}
\label{WZSWV}
    W_{mag} = M^2 B
\end{equation}

In the rest of this section we provide the physical derivation of this confining duality.
We start by adding a singlet $\alpha$ in the electric theory, flipping the meson $M$ through a superpotential 
\begin{equation}
\Delta W_{ele} = \alpha Q_1 Q_2
\end{equation}
In this way the dual superpotential becomes 
\begin{equation}
\label{WZSWVflipped}
W_{mag} = \alpha M + M B^2
\end{equation}
that vanishes once we compute the F-terms of the massive fields $\alpha$ and $M$.
The next step consists in deconfining the field $A$. We distinguish two cases, depending on the parity of $N_c$.
Let us study them separately.

\subsection{Deconfinement with odd $N_c=2k+1$}

\begin{figure}[H]
    \centering
    \begin{tikzpicture}[auto, scale=1.2]
        %%%%%%%%%%%%%% nodes %%%%%%%%%%
        \node [circle, draw=red!50, fill=red!20, inner sep=2pt, minimum size=5mm] (N) at (0,0) {$2k+2$};
        \node [rectangle, draw=red!50, fill=red!20, inner sep=2pt, minimum size=4mm] (FL) at (-2.5,0) {$2k+4$};
        \node [rectangle, draw=blue!50, fill=blue!20, inner sep=2pt, minimum size=4mm] (FR) at (2,0) {$4k+2$};
        \node [circle, draw=blue!50, fill=blue!20, inner sep=2pt, minimum size=5mm] (M) at (0,-2) {$2k$};
        \node [rectangle, draw=red!50, fill=red!20, inner sep=2pt, minimum size=4mm] (FD) at (-2.5,-2) {$2$};
    
        %%%%%%%%%%%%% fields %%%%%%%%%%
        \begin{scope}[on background layer,decoration={
            markings,
            mark=at position 0.5 with {\arrow{>}}}
            ]
            \draw [postaction={decorate}, thick, above] (M) [] to node {$L$} (FD);
            \draw [postaction={decorate}, thick] (N) to node {$Q_1$} (FL);
            \draw [postaction={decorate}, thick] (N) to node {$B$} (M); 
            \draw [postaction={decorate}, thick] (FR) to node {$Q_2$} (N);
            \draw [postaction={decorate}, thick] (FD) to node {$C$} (N);
            \draw [postaction={decorate}, thick, bend left=60] (FL) to node {$\alpha$} (FR);
            \draw (FD) to [out=135, in=225, looseness=6] (FD) [<-,thick];
        \end{scope}
    
        \node [left=0.5cm of FD] (L) {$\beta$};
    \end{tikzpicture}
    \caption{Quiver description of the model after the deconfining of  the $\SU{2k+2}$ rank-$2$ conjugate anti-symmetric tensor A. Gauge groups are represented as circles while flavor nodes are represented with squares. Symplectic groups are depicted in blue and unitary groups are depicted in red.}
    \label{1stepdecodd}
\end{figure}
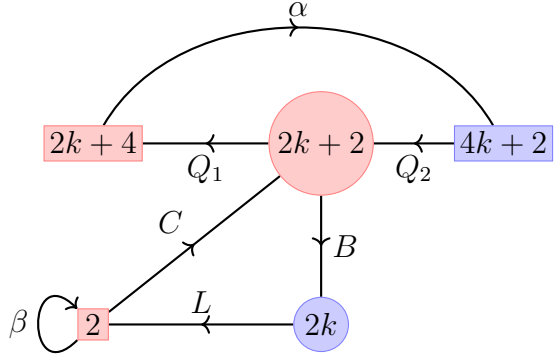

In this case we deconfine the rank-$2$ conjugate anti-symmetric tensor $A$ of 
$\SU{2k+2}$ 
We depicted the model in Figure \ref{1stepdecodd} in terms of a quiver gauge theory.
The superpotential for the deconfined model is schematically
\begin{equation}
\label{Wdec1}
W= (B Q_2)^2  + \alpha Q_1 Q_2 + BLC+\beta L^2
\end{equation} 
where the field $\beta$ corresponds to the $2 \times 2$ antisymmetric matrix, i.e. it is a singlet.
The anti-symmetric tensor $A$ is recovered by confining the $\USp{2k}$ gauge node in terms of the field B, i.e. $A \sim B^2$, where the contraction is done on the $\USp{2k}$ indices.

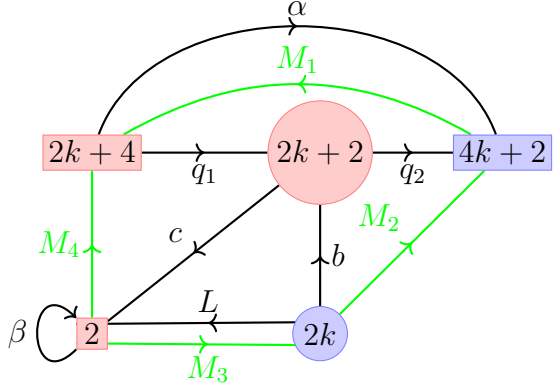
\begin{figure}[H]
    \centering
    \begin{tikzpicture}[auto, scale=1.2]
        %%%%%%%%%%%%%% nodes %%%%%%%%%%
        \node [circle, draw=red!50, fill=red!20, inner sep=2pt, minimum size=5mm] (N) at (0,0) {$2k+2$};
        \node [rectangle, draw=red!50, fill=red!20, inner sep=2pt, minimum size=4mm] (FL) at (-2.5,0) {$2k+4$};
        \node [rectangle, draw=blue!50, fill=blue!20, inner sep=2pt, minimum size=4mm] (FR) at (2,0) {$4k+2$};
        \node [circle, draw=blue!50, fill=blue!20, inner sep=2pt, minimum size=5mm] (M) at (0,-2) {$2k$};
        \node [rectangle, draw=red!50, fill=red!20, inner sep=2pt, minimum size=4mm] (FD) at (-2.5,-2) {$2$};
    
        %%%%%%%%%%%%% fields %%%%%%%%%%
        \begin{scope}[on background layer,decoration={
            markings,
            mark=at position 0.5 with {\arrow{>}}}
            ]
            \draw [postaction={decorate}, thick,above] ($(M.north east)!0.2!(M.south east)$) [] to node {$L$} ($(FD.north west)!0.2!(FD.south west)$);
            \draw [postaction={decorate}, thick, below,color=green] ($(FD.south west)!0.2!(FD.north west)$) [] to node {$M_3$} ($(M.south east)!0.2!(M.north east)$);
            \draw [postaction={decorate}, thick, color=green] (FD) to node {$M_4$} (FL);
            \draw [postaction={decorate}, thick, below] (FL) to node {$q_1$} (N);
            \draw [postaction={decorate}, thick, right] (M) to node {$b$} (N); 
            \draw [postaction={decorate}, thick, below] (N) to node {$q_2$} (FR);
            \draw [postaction={decorate}, thick, color=green] (M) to node {$M_2$} (FR);
            \draw [postaction={decorate}, thick, above left] (N) to node {$c$} (FD);
            \draw [postaction={decorate}, thick, bend right=30, color=green, above] (FR) to node {$M_1$} (FL);
            \draw [postaction={decorate}, thick, bend left=70] (FL) to node {$\alpha$} (FR);
            \draw (FD) to [out=135, in=225, looseness=6] (FD) [<-,thick];
        \end{scope}
    
        \node [left=0.5cm of FD] (L) {$\beta$};
    \end{tikzpicture}
    \caption{Quiver obtained after Seiberg duality on $\SU{2k+2}$. The rank of the 
            dual gauge group is the same as above, but there are new mesonic degresse of freedom that modify the superpotential.}
    \label{2stepdualodd}
\end{figure}

The next step consists of Seiberg duality on $\SU{2k+2}$. This gauge group is self dual and the quiver is represented in Figure \ref{2stepdualodd}.
The superpotential of the dual theory is
\begin{equation}
\label{Wdualodd}
    W = (M_2)^2 + \alpha M_1  + M_3L + \beta L^2 + M_1q_1q_2 + M_2bq_2 + M_3bc + M_4q_1c 
\end{equation}
where $b,c,q_1,q_2$ are the dual quarks and the mesons $M_{1,2,3,4}$ are associated to the quarks of the previous phase through the dictionary
\begin{equation}
\label{mapmes}
   M_1 \longleftrightarrow Q_1 Q_2, \quad
         M_2 \longleftrightarrow B Q_2, \quad 
         M_3 \longleftrightarrow BC,  \quad
         M_4 \longleftrightarrow Q_1C
\end{equation}

By integrating out the massive fields this superpotential becomes
\begin{equation}
\label{Wdualodd2}
W  = (bq_2)^2 + \beta (bc)^2 + M_4 q_1 c 
\end{equation}

\begin{figure}[H]
    \centering
    \begin{tikzpicture}[auto, scale=1.2]
        %%%%%%%%%%%%%% nodes %%%%%%%%%%
        \node [circle, draw=red!50, fill=red!20, inner sep=2pt, minimum size=5mm] (N) at (0,0) {$2k+2$};
        \node [rectangle, draw=red!50, fill=red!20, inner sep=2pt, minimum size=4mm] (FL) at (-2.5,0) {$2k+4$};
        \node [rectangle, draw=blue!50, fill=blue!20, inner sep=2pt, minimum size=4mm] (FR) at (2,0) {$4k+2$};
        \node [rectangle, draw=red!50, fill=red!20, inner sep=2pt, minimum size=4mm] (FD) at (-2.5,-2) {$2$};
    
        %%%%%%%%%%%%% fields %%%%%%%%%%
        \begin{scope}[on background layer,decoration={
            markings,
            mark=at position 0.5 with {\arrow{>}}}
            ]
            \draw [postaction={decorate}, thick, color=green] (FD) to node {$M_4$} (FL);
            \draw [postaction={decorate}, thick, below] (FL) to node {$q_1$} (N);
            \draw [postaction={decorate}, thick, below] (N) to node {$q_2$} (FR);
            \draw [postaction={decorate}, thick, above left] (N) to node {$c$} (FD);
            \draw (FD) to [out=135, in=225, looseness=6] (FD) [<-,thick];
            \draw (N) to [out=70, in=110, looseness=6] (N) [<-,thick];
        \end{scope}
    
        \node [left=0.5cm of FD] (L) {$\beta$};
        \node [above=0.7cm of N] (N) {$a$};
    \end{tikzpicture}
    \caption{Quiver obtained after confining the $\USp{2k}$ gauge node. The anti-symmetric field $a$ gets a vev from the quantum constraint on the moduli space.}
    \label{3stepcsb}
\end{figure}
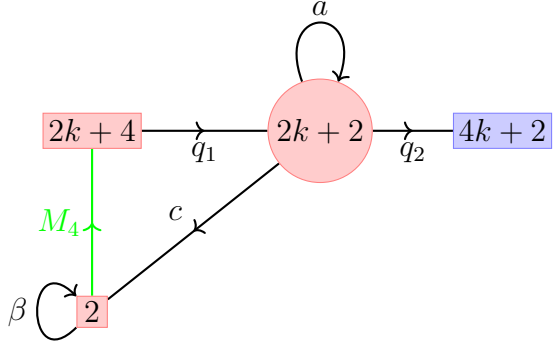

Then we consider the $\USp{2k}$ gauge group with $2k+2$ flavors. This gauge theory confines 
with a quantum constraint enforced on the moduli space. By considering the low energy dynamics we are left with a single gauge group $\SU{2k+2}$ and the field content can be read from the quiver in Figure \ref{3stepcsb}.

The superpotential becomes
\begin{equation}
\label{WchiSB}
W = aq_2^2 + \beta a c^2 + M_4 q_1 c + \lambda (\text{Pf}(a) - \Lambda^{2k+2}) 
\end{equation}
The last term in (\ref{WchiSB}) enforces the quantum constraint on the moduli space through the Lagrange multiplier $\lambda$. \\
The constraint breaks the gauge symmetry to $\USp{2(k+1)}$ and this Higgsing gives mass to $Q_2$ as well.
The leftover superpotential is 
\begin{equation}
    \label{WchiSB2}
    W =  \beta c^2 + M_4 q_1 c 
\end{equation}
This model s-confines and the final superpotential is
\begin{equation}
    \label{lastsconf}
    W = \beta  \gamma + M_4 L +\text{Pf} \left(
    \begin{array}{cc}
    \gamma & L \\
    -L^T&U 
    \end{array}\right)
\end{equation}
where $L,U$ and $\gamma$ correspond to the $\USp{2k+2}$ contractions
$q_1 c$, $q_1^2$ and $c^2$ respectively.
Integrating out the massive fields we are left with just the  
meson $L$ in the anti-symmetric representation of the flavor symmetry group 
$\SU{N_c+3} = \SU{2k+4}$. This field corresponds indeed to the baryon $B$ expected in the SWV duality.

In order to connect with the WZ superpotential of the  SWV duality (\ref{WZSWV}) we have to flip the field $\alpha$ in (\ref{WZSWV}). 
This turns off the field $\alpha$ in the derivation and keeps the meson $M_1$ massless in (\ref{Wdualodd}).
The superpotential (\ref{Wdualodd2}) then becomes 
\begin{equation}
\label{Wdualodd2bis}
W  = (bq_2)^2 + \beta (bc)^2 + M_4 q_1 c +M_1 q_1 q_2
\end{equation}
The other steps in the derivation are straightforward and in the final superpotential
(\ref{lastsconf}) there is a further contribution $\Delta W \propto M_1^2 U$.
This terms survives after the massive fields are integrated out and by the identification $U \leftrightarrow B$ and $M_1 \leftrightarrow M$ we obtain exactly 
(\ref{WZSWV}) as expected.

This concludes the proof of the duality from the field theory analysis in the case with $N_c=2k+1$.
Before moving to $N_c=2k$ it is instructive to reproduce the analysis using the $S^3 \times S^1$ supersymmetric index.

In order to have a better physical intuition of the duality from localization 
we start by  rewriting $I_E$ and $I_M$ by modifying the $\USp{2N_c}$ fugacities as
$t \rightarrow t \sqrt{pq/S}$.
This gives 
\begin{equation}
\label{newele}
\begin{split}
    I_E &= \frac{(p,p)^{N_c}_{\infty} (q,q)^{N_c}_{\infty}}{(N_c+1)!} \int_{\mathbb{T}^{N_c}} \prod_{1 \leq i < j \leq N_c+1}       \frac{\Gamma (S z_i^{-1} z_j^{-1}; p,q)}{\Gamma(z_i^{-1} z_j, z_i^{-1}z_j; p,q)} \\
    &\times \prod_{j=1}^{N_c+1} \prod_{m=1}^{N_c+3} \Gamma(s_m z_j^{-1}; p,q) \prod_{k=1}^{N_c} \Gamma( \sqrt{\frac{pq}{S}} t_k^{\pm 1} z_j; p,q) \prod_{j=1}^{N_c} \frac{\dd{z_j}}{2 \pi i z_j},
\end{split}
\end{equation}
and
\begin{equation}
\label{newmag}
    I_M = \prod_{k=1}^{N_c} \prod _{m=1}^{N_c+3} \Gamma (s_m t_k^{\pm 1} \sqrt{\frac{pq}{S}}; p,q) \prod_{1 \leq l < m \leq N_c+3} \Gamma(S s_l^{-1} s_m^{-1}; p,q),
\end{equation}
again with the  balancing condition  $S = \prod_{m=1}^{N_c+3} s_m$.

\begin{figure}[H]
    \centering
    \begin{tikzpicture}[auto, scale=2]
        %%%%%%%%%%%%%% nodes %%%%%%%%%%
        \node [circle, draw=red!50, fill=red!20, inner sep=2pt, minimum size=5mm] (N) at (0,0) {$2k+2$};
        \node [rectangle, draw=red!50, fill=red!20, inner sep=2pt, minimum size=4mm] (FL) at (-2.5,0) {$2k+4$};
        \node [rectangle, draw=blue!50, fill=blue!20, inner sep=2pt, minimum size=4mm] (FR) at (2,0) {$4k+2$};
        \node [circle, draw=blue!50, fill=blue!20, inner sep=2pt, minimum size=5mm] (M) at (0,-2) {$2k$};
        \node [rectangle, draw=red!50, fill=red!20, inner sep=2pt, minimum size=4mm] (FD) at (-2.5,-2) {$2$};
    
        %%%%%%%%%%%%% fields %%%%%%%%%%
        \begin{scope}[on background layer,decoration={
            markings,
            mark=at position 0.5 with {\arrow{>}}}
            ]
            \draw [postaction={decorate}, thick, color=green] (FD) to node {$M_4$} (FL);
            \draw [postaction={decorate}, thick, below] (FL) to node {$q_1$} (N);
            \draw [postaction={decorate}, thick, left] (M) to node {$b$} (N); 
            \draw [postaction={decorate}, thick, below] (N) to node {$q_2$} (FR);
            \draw [postaction={decorate}, thick, above left] (N) to node {$c$} (FD);
            \draw (FD) to [out=135, in=225, looseness=6] (FD) [<-,thick];
        \end{scope}
    
        \node [left=0.5cm of FD] (L) {$\beta$};
        \node (q2) at (1,.2) {{\color{blue}$\sqrt{pq}z_i^{-1}t_k^{\pm1}$}};
        \node (q1) at (-1.3,.2) {{\color{blue}$\sqrt{S}s_a^{-1}z_i$}};
        \node (b) at (.4,-1.1) {{\color{blue}$z_i x_u^{\pm 1}$}};
        \node [rotate=40] (c) at (-.8,-1.1) {{\color{blue}$\sqrt{\dfrac{pq}{S^{k+1}}}z_i^{-1}y_m$}};
        \node (M4) at (-1.9,-.6) {{\color{blue}$\sqrt{pq S^{k}}s_a y_m^{-1}$}};
    \end{tikzpicture}
    \caption{$\SU{2k+4}\times \USp{2k}$ quiver before confining the symplectic node.}
    \label{aftermasses}
\end{figure}
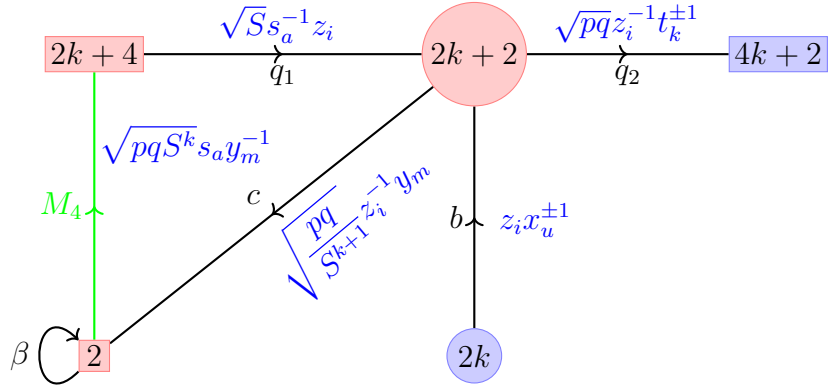

Then we proceed to deconfine the rank-$2$ conjugate anti-symmetric tensors, to dualize the $\SU{2k+4}$ node and  to
integrate out the massive fields. These  steps are done by using the integral identities collected in \cite{Spiridonov:2009za} (which are reproduced in the appendix \ref{appendixA})  and the reflection equation  for the elliptic gamma functions $\Gamma_{e}(pq /x) \Gamma_e(x)=1$.
We skip these standard elementary steps and focus on the quiver described in Figure \ref{aftermasses}, where we also highlighted in blue the fugacity of each field in the 
$S^3 \times S^1$ supersymmetric index.

The $S^3 \times S^1$ supersymmetric index for these models is given by formula
\begin{equation}
    \begin{split}
         I_E =& 
          \frac{(p,p)^{2k+1}_{\infty} (q,q)^{3k+1}_{\infty}}{2^k k! (3k+2)!} 
         \Gamma (S^{k+1}; p,q) \prod_{a=1}^{2k+4} \prod_{m=1,2} \Gamma (\sqrt{pqS^k} s_a y_m^{-1}) 
          \\
         \times&
         \int_{\mathbb{T}^{2k-1}}  \int_{\mathbb{T}^{k}} 
          \prod_{u=1}^{k} \frac{\dd{x_u}}{2 \pi i x_u} \prod_{i=1}^{2k+2} \frac{\dd{z_i}}{2 \pi i z_i}
         \prod_{1 \leq u < v \leq k} \frac{1}{\Gamma(x_u^{\pm 1} x_v^{\pm 1}; p,q)} \prod_{u=1}^{k} \frac{\prod_{i=1}^{2k+2} \Gamma(z_i x_u^{\pm 1}; p,q)}{\Gamma(x_u^{\pm 2}; p,q)}
         \\
         \times&\frac{\prod_{i=1}^{2k+2} \prod_{a=1}^{2k+4} \Gamma (\sqrt{S} s_a^{-1} z_i) \prod_{m=1,2} \Gamma (\sqrt{pq/S^{k+1}} z_i^{-1} y_m) \prod_{k=1}^{2k +1} \Gamma (\sqrt{pq} z_i^{-1} t_k^{\pm 1})}{\prod_{1 \leq i < j \leq 2k+2} \Gamma (z_i z_j^{-1}, z_i^{-1} z_j; p,q)} 
    \end{split}
    \label{eq_I_E}
\end{equation}

We then consider the change of variables  $z_i = e^{ 2 \pi i \phi_i}$, where  $\phi_i$ are real and the balancing condition is  $\sum_{i=1}^{2k+2} \phi_i = 0$. With such a change of variables we can substitute in the index the following terms 
\begin{equation}
    \begin{split}
        &\frac{(p,p)^{k}_{\infty} (q,q)^{k}_{\infty}}{2^k k!} \int_{\mathbb{T}^{k}} \prod_{1 \leq u < v \leq k} \frac{1}{\Gamma(x_u^{\pm 1} x_v^{\pm 1}; p,q)} \prod_{u=1}^{k} \frac{\prod_{i=1}^{2k+2} \Gamma(e^{2 \pi i \phi_i} x_u^{\pm 1}; p,q)}{\Gamma(x_u^{\pm 2}; p,q)} \prod_{u=1}^{k} \frac{\dd{x_u}}{2 \pi i x_u}  \\
        & = \frac{1}{(p;p)_{\infty}^{k} (q,q)^{k}_{\infty} } \sum_{(\Phi_1 \bigcup \Phi_2)/ S_2^k} \prod_{1 \leq i < j \leq k+1} \Gamma(e^{2 \pi i (\pm \tilde {\phi}_i \pm \tilde {\phi}_j )};p,q) \sum_{S_{k+1} (\Phi_2)} \prod_{i=1}^{k} \delta (\Tilde{\phi}_i + \Tilde{\phi}_{k+1+i} ),
     \end{split}
\end{equation}
where we used the identity (\ref{A6}). This identity was derived in \cite{Spiridonov:2014cxa} and it represents the evaluation of the superconformal index for $\USp{2M}$ SQCD with $2M+2$ fundamentals. The fact that the models confines with a quantum  superpotential that breaks the chiral symmetry is reflected in the structure of the 
$\delta$-functions in (\ref{A6}). 
 In this way (\ref{eq_I_E}) becomes
\begin{equation}
\label{Indexcsb}
    \begin{split}
        &I_E = \Gamma (S^{k+1}; p,q) \prod_{a=1}^{2k+4} \prod_{m=1,2} \Gamma (\sqrt{pqS^k} s_a y_m^{-1})
                 \frac{(p,p)^{k+1}_{\infty} (q,q)^{k+1}_{\infty}}{(2k+2)!} \int_{\mathbb{T}^{2k-1}} 
         \prod_{i=1}^{2k+2} \frac{\dd{z_i}}{2 \pi i z_i}
          \\
         & \frac{\prod_{i=1}^{2k+2} \prod_{a=1}^{2k+4} \Gamma (e^{2 \pi i \phi_i}\frac{\sqrt{S}}{s_a} ) \prod_{m=1,2} \Gamma (\sqrt{\frac{pq}{S^{k+1}}} e^{- 2 \pi i \phi_i} y_m) \prod_{k=1}^{2k +1} \Gamma (\sqrt{pq} e^{- 2 \pi i \phi_i} t_k^{\pm 1})}{\prod_{1 \leq i < j \leq 2k+2} \Gamma (e^{ 2 \pi i (\phi_i - \phi_j)}, e^{ 2 \pi i (-\phi_i + \phi_j)}; p,q)} \\
         &  \sum_{(\Phi_1 \bigcup \Phi_2)/ S_2^k} \prod_{1 \leq i < j \leq k+1} \Gamma(e^{2 \pi i (\pm \tilde {\phi}_i \pm \tilde {\phi}_j )};p,q) \sum_{S_{k+1} (\Phi_2)} \prod_{i=1}^{k} \delta (\Tilde{\phi}_i + \Tilde{\phi}_{k+1+i} ), 
    \end{split}
\end{equation}
where  $\Phi_1 = (\tilde{\phi}_1, ... , \tilde{\phi}_k, \tilde{\phi}_{k+1} = \phi_{k+1})$ and $\Phi_2 = (\tilde{\phi}_{k+2}, ... , \tilde{\phi}_{2k+2})$.
Using the constraints imposed by the balancing condition $\sum_{i=1}^{2k+2} \phi_i = 0$, the delta functions and the reflection equation we can simplify (\ref{Indexcsb}) to
\begin{equation}
    \begin{split}
        I_E = & \Gamma (S^{k+1}; p,q) \prod_{a=1}^{2k+4} \prod_{m=1,2} \Gamma (\sqrt{pqS^k} s_a y_m^{-1})
        \frac{(p,p)^{k+1}_{\infty} (q,q)^{k+1}_{\infty}}{(k+1)! \cdot 2^{k+1}} \int_{\mathbb{T}} 
        \prod_{i=1}^{k+1} \frac{\dd{z_i}}{2 \pi i z_i}
       \\
        & \frac{\prod_{i=1}^{k+1} \prod_{a=1}^{2k+4} \Gamma (\sqrt{S} s_a^{-1} z_i^{\pm 1}) \prod_{m=1,2} \Gamma (\sqrt{pq/S^{k+1}} z_i^{\pm 1} y_m)}{\prod_{1 \leq i < j \leq k+1} \Gamma (z_i^{\pm 1} z_j^{\pm 1}; p,q) \prod_{i=1}^{k+1} \Gamma (z_i^{\pm 2}; p,q)},
    \end{split}
\end{equation}
that represents the s-confining $\USp{2k+2}$ theory with $2k+6$ fundamentals and superpotential (\ref{WchiSB}).
Using the limiting case identity associated to this confining duality (i.e. formula (\ref{uspid})
for $N_f=N_c+4$) the identity between (\ref{newele}) and  (\ref{newmag})
 is then correctly recovered. This concludes the proof of the derivation of the identity of \cite{spiridonov2004inversions} from the physical approach when $N_c=2k+1$ 

\subsection{Deconfinement with even $N_c=2k$}

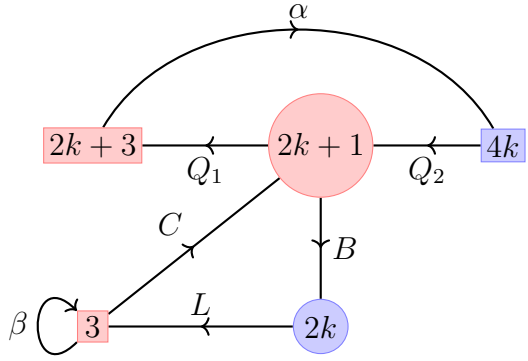
\begin{figure}[H]
    \centering
    \begin{tikzpicture}[auto, scale=1.2]
        %%%%%%%%%%%%%% nodes %%%%%%%%%%
        \node [circle, draw=red!50, fill=red!20, inner sep=2pt, minimum size=5mm] (N) at (0,0) {$2k+1$};
        \node [rectangle, draw=red!50, fill=red!20, inner sep=2pt, minimum size=4mm] (FL) at (-2.5,0) {$2k+3$};
        \node [rectangle, draw=blue!50, fill=blue!20, inner sep=2pt, minimum size=4mm] (FR) at (2,0) {$4k$};
        \node [circle, draw=blue!50, fill=blue!20, inner sep=2pt, minimum size=5mm] (M) at (0,-2) {$2k$};
        \node [rectangle, draw=red!50, fill=red!20, inner sep=2pt, minimum size=4mm] (FD) at (-2.5,-2) {$3$};
    
        %%%%%%%%%%%%% fields %%%%%%%%%%
        \begin{scope}[on background layer,decoration={
            markings,
            mark=at position 0.5 with {\arrow{>}}}
            ]
            \draw [postaction={decorate}, thick, above] (M) [] to node {$L$} (FD);
            \draw [postaction={decorate}, thick] (N) to node {$Q_1$} (FL);
            \draw [postaction={decorate}, thick] (N) to node {$B$} (M); 
            \draw [postaction={decorate}, thick] (FR) to node {$Q_2$} (N);
            \draw [postaction={decorate}, thick] (FD) to node {$C$} (N);
            \draw [postaction={decorate}, thick, bend left=60] (FL) to node {$\alpha$} (FR);
            \draw (FD) to [out=135, in=225, looseness=6] (FD) [<-,thick];
        \end{scope}
    
        \node [left=0.5cm of FD] (L) {$\beta$};
    \end{tikzpicture}
    \caption{Quiver description of the model after the deconfining of  the $\SU{2k+1}$ rank-$2$ conjugate anti-symmetric tensor A}
    \label{1stepdeceven}
\end{figure}

In this case we deconfine the rank-$2$ conjugate anti-symmetric tensor $A$ of 
$\SU{2k+1}$ 
We depicted the model in Figure \ref{1stepdeceven} in terms of a quiver gauge theory.

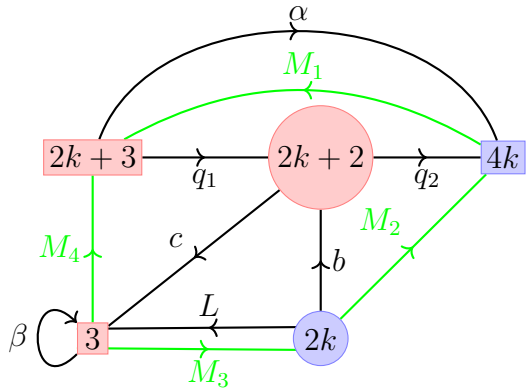
\begin{figure}[H]
\centering
\begin{tikzpicture}[auto, scale=1.2]
    %%%%%%%%%%%%%% nodes %%%%%%%%%%
    \node [circle, draw=red!50, fill=red!20, inner sep=2pt, minimum size=5mm] (N) at (0,0) {$2k+2$};
    \node [rectangle, draw=red!50, fill=red!20, inner sep=2pt, minimum size=4mm] (FL) at (-2.5,0) {$2k+3$};
    \node [rectangle, draw=blue!50, fill=blue!20, inner sep=2pt, minimum size=4mm] (FR) at (2,0) {$4k$};
    \node [circle, draw=blue!50, fill=blue!20, inner sep=2pt, minimum size=5mm] (M) at (0,-2) {$2k$};
    \node [rectangle, draw=red!50, fill=red!20, inner sep=2pt, minimum size=4mm] (FD) at (-2.5,-2) {$3$};

    %%%%%%%%%%%%% fields %%%%%%%%%%
    \begin{scope}[on background layer,decoration={
        markings,
        mark=at position 0.5 with {\arrow{>}}}
        ]
        \draw [postaction={decorate}, thick,above] ($(M.north east)!0.2!(M.south east)$) [] to node {$L$} ($(FD.north west)!0.2!(FD.south west)$);
        \draw [postaction={decorate}, thick, below,color=green] ($(FD.south west)!0.2!(FD.north west)$) [] to node {$M_3$} ($(M.south east)!0.2!(M.north east)$);
        \draw [postaction={decorate}, thick, color=green] (FD) to node {$M_4$} (FL);
        \draw [postaction={decorate}, thick, below] (FL) to node {$q_1$} (N);
        \draw [postaction={decorate}, thick, right] (M) to node {$b$} (N); 
        \draw [postaction={decorate}, thick, below] (N) to node {$q_2$} (FR);
        \draw [postaction={decorate}, thick, color=green] (M) to node {$M_2$} (FR);
        \draw [postaction={decorate}, thick, above left] (N) to node {$c$} (FD);
        \draw [postaction={decorate}, thick, bend right=30, color=green, above] (FR) to node {$M_1$} (FL);
        \draw [postaction={decorate}, thick, bend left=70] (FL) to node {$\alpha$} (FR);
        \draw (FD) to [out=135, in=225, looseness=6] (FD) [<-,thick];
    \end{scope}

    \node [left=0.5cm of FD] (L) {$\beta$};
\end{tikzpicture}
\caption{Quiver obtained after Seiberg duality on $\SU{2k+1}$. Curiously in this case the dual gauge group increases its rank becoming $\SU{2k+2}$.}
\label{2stepdualeven}
\end{figure}

The superpotential for the deconfined model is given again by formula (\ref{Wdec1}).
The next step consists of Seiberg duality on $\SU{2k+1}$. The dual gauge group is $\SU{2k+2}$ and the quiver is represented in Figure \ref{2stepdualeven}.
The superpotential of the dual theory is again given by  (\ref{Wdualodd}).

The fields $b,c,q_1,q_2$ are the dual quarks and the mesons $M_{1,2,3,4}$ are associated to the quarks of the previous phase through the dictionary spelled out in (\ref{mapmes}).
By integrating out the massive fields the superpotential becomes the one in (\ref{Wdualodd2}).
Then we observe that $\USp{2k}$ with $2k+2$ flavors confines with a quantum moduli space and after such confinement we are left with a $\SU{2k+2}$ gauge group with superpotential 
(\ref{WchiSB}). The partial Higgsing triggered by the quantum constraint enforced by the Lagrange multiplier reduces the theory to $\USp{2k+2}$ with $2k+6$ fundamentals and superpotential
(\ref{lastsconf}). Integrating out the massive fields we are left with a single field $L$ that 
correspond to the baryon $B$ of the original theory.

The analysis for even $N_c$ is then almost identical to the case of odd $N_c$. 
For this reason we skip the derivation of the duality from the superconformal index. 
The interested reader can reproduce it by following the stepwise procedure that we described for odd $N_c$.

\section{The Okazaki-Smith 3d duality}

In this section we study a 3d $\mathcal{N}=2$ confining duality recently proposed in \cite{Okazaki:2023hiv}.
The electric model is $\USp{4}$ SQCD with two fundamentals and two rank-$2$ anti-symmetric tensors.
The model has an $\U{2}^2 \times \U{1}_R$ global symmetry and the charges of the fields
under these symmetries are summarized in (\ref{OScharges}). 
\begin{equation}
\label{OScharges}
\begin{array} {c|c||ccccc}
& \USp{4} & \SU{2}_A & \SU{2}_a& \U{1}_A& \U{1}_a& U(1)_R\\
\hline
A & 6 &2&1&1&0&0 \\
Q & 4 &1&2&0&1&0
\end{array}
\end{equation}

The model has vanishing superpotential and its low energy dynamics is described by the gauge invariant combinations
$M=Q_1 Q_2$, $\phi_I = \Tr A_I$, $\phi_{IJ} = \Tr (A_I A_J)$, $B_{\alpha \beta} = Q_\alpha A_1 A_2 Q_{\beta}$ and $B_I = Q_1 \phi_I Q_2$. These fields interact through a superpotential with a singlet $\mathcal{T}_4$ that corresponds to  the minimal monopole of $\USp{4}$. 
The charges of the fields with respect to the global $\U{2}^2 \times \U{1}_R$ symmetry are:
\begin{equation}
\label{OScharges2}
\begin{array} {c|ccccc}
& \SU{2}_A & \SU{2}_a& \U{1}_A& \U{1}_a& U(1)_R\\
\hline
M  & 1& 1&  0& 2& 0\\
B_{\alpha \beta}  &1 &3 &2 &2 &0  \\
\phi_{IJ} &3 &1 &2 &0   &0\\
\phi_{I}  &2 &1 &1 &0 &0 \\
B_{I}   & 2& 1&1 &2   &0\\
\mathcal{T}_4    & 1&1 & -4&-4&2  \\
\end{array}
\end{equation}

In the following we will derive this confining duality by deconfining the anti-symmetric tensors and then by sequentially dualizing the gauge groups.
We found that in order to proceed it is very useful to add to the electric theory an $\SU{2}_A$ vector $\vec s=(s_1,s_2)$ interacting with $\Pf{A_1}$ and $\Pf{A_2}$ through the superpotential 
\begin{equation}
\label{Wflip}
    W = \vec s \cdot \Pf{\vec A} =\sum_{I=1,2}s_{I}\Pf{A_{I}}=\frac{1}{8}\sum_{I=1,2}s_I\big(\Tr(A_I)^2-2\Tr(A_I^2)\big)
\end{equation}
where the trace of an anti-symmetric matrix is defined as $\Tr A_I = A_I^{ij} J_{ij}$.

\subsection{Field theory analysis}

In the following we will derive the duality using the field theory approach.
We proceed by representing the model in terms of a quiver gauge theory, using the same conventions of the previous section: the blue circles refer to symplectic gauge groups while the red squares identify the special unitary flavor groups.

We start by  considering the model with the flip in formula (\ref{Wflip})
\begin{equation}
\centering
\begin{tikzpicture}[auto,baseline=($(F.base)!.5!(N.base)$)]
	\node [circle, draw=blue!50, fill=blue!20, inner sep=0pt, minimum size=5mm] (N) at (0,0) {$4$};
	\node[rectangle, draw=red!50, fill=red!20, inner sep=0pt, minimum size=4mm] (F) at (0,-1.5) {$2$};
	
	\draw (N) to node {$Q$} (F) [-,thick];
	\draw (N) to [out=180, in=120, looseness=8] (N) [-,thick];
	\draw (N) to [out=0, in=60, looseness=8] (N) [-, thick];
	
	\node [above left=0.4cm of N] (M) {$A_{1}$};
	\node [above right=0.4cm of N] (M) {$A_{2}$};
\end{tikzpicture}\qquad
 	W=\sum_{I=1,2}s_{I}\Pf{A_{I}}
\end{equation}
We  then deconfine the two rank-$2$ anti-symmetric fields $A_{I}$ with two auxiliary  $\USp{2}$ nodes with the assignment 
\begin{equation}
	A_{1}^{ij}=q_{1}^{\alpha_1 i}q_{1}^{\beta_1 j}\epsilon_{\alpha_1 \beta_1 },\qquad A_{2}^{ij}=q_{2}^{\alpha_2\,i}q_{2}^{\beta_2\,j}\epsilon_{\alpha_2 \beta_2}
\end{equation}
where the $i$-index refers to the $\USp{4}$ node and $(\alpha_{1,2},\beta_{1,2})$ are indices of the two $\USp{2}_{1,2}$ gauge groups.  Therefore the deconfined theory is
\begin{equation}
\label{decft}
\centering
\begin{tikzpicture}[auto,baseline=(N)]
	\node [circle, draw=blue!50, fill=blue!20, inner sep=0pt, minimum size=5mm] (N) at (0,0) {$4$};
	\node[rectangle, draw=red!50, fill=red!20, inner sep=0pt, minimum size=4mm] (F) at (0,-1.5) {$2$};
	\node [circle, draw=blue!50, fill=blue!20, inner sep=0pt, minimum size=5mm] (A1) at (-1.2,1.2) {$2$};
	\node [circle, draw=blue!50, fill=blue!20, inner sep=0pt, minimum size=5mm] (A2) at (1.2,1.2) {$2$};
	
	\draw (N) to node {$Q$} (F) [-,thick];
	\draw (N) to node {$q_{1}$} (A1) [-,thick];
	\draw (N) to node [below right] {$q_{2}$} (A2) [-, thick];
\end{tikzpicture}\qquad
 	W=0
	%t_I P_I
\end{equation}
Observe that the superpotential is vanishing because the singlets $s_{1,2}$ have flipped the monopoles of  the $\USp{2}_{1,2}$ gauge groups. 
The central $\USp{4}$ node in this theory has then $6$ fundamentals and therefore it confines \cite{Aharony:1997gp}. The IR description has then two $\USp{2}_{1}$ and $\USp{2}_{2}$ gauge groups connected by a bifundamental field. There is still a manifest $\SU{2}$ flavor  symmetry associated to a node in the quiver and there are further fundamental fields for both the $\USp{2}_{1,2}$ gauge factors.
There is also a singlet $Y_4$ identified with the monopole of the $\USp{4}$ gauge group for the model in (\ref{decft}), that interacts through a superpotential with the generalized meson of $\USp{4}$  itself.
The quiver and the superpotential for this dual theory are
\vspace{\belowdisplayskip}

\begin{minipage}[c]{0.3\linewidth}
\centering
    \begin{tikzpicture}[auto,baseline=(A1)]
        %%%%%%%%%%%%%% nodes %%%%%%%%%%
        \node[rectangle, draw=red!50, fill=red!20, inner sep=0pt, minimum size=4mm] (F) at (0,-1) {$2$};
        \node [circle, draw=blue!50, fill=blue!20, inner sep=0pt, minimum size=5mm] (A1) at (-1.2,1.2) {$2$};
        \node [circle, draw=blue!50, fill=blue!20, inner sep=0pt, minimum size=5mm] (A2) at (1.2,1.2) {$2$};
        
        %%%%%%%%%%%%% fields %%%%%%%%%%
        \draw (A1) [bend right=30] to node [left=.2cm] {$X_{13}$} (F) [-, thick];
        \draw (A2) [bend left=30] to node [right=.2cm] {$X_{23}$} (F) [-, thick];
        \draw (A1) [bend left=30] to node {$X_{12}$} (A2) [-,thick];
        \draw (A1) to [out=180, in=120, looseness=8] (A1) [-,thick];
        \draw (A2) to [out=0, in=60, looseness=8] (A2) [-, thick];
        \draw (F) to [out=-45, in=-135, looseness=6] (F) [-, thick];
            
        \node [above left=0.4cm of A1] (X11) {$X_{11}$};
        \node [above right=0.4cm of A2] (X22) {$X_{22}$};
        \node [below=0.5cm of F] (X33) {$X_{33}$};
        \node [below right=.1 pt of A1] (A) {$\scriptstyle \alpha_1$};
        \node [below left=.1 pt of A2] (a) {$\scriptstyle  \alpha_2$};
        \node [above=.1 pt of F] (alpha) {$\scriptstyle\alpha$};
    \end{tikzpicture}
\end{minipage}\hfill
\begin{minipage}[c]{0.5\linewidth}
\begin{equation}
\label{dualusp4ft}
\begin{split}
	W&=Y_{4}\Pf{X}\\
	&=Y_{4}\epsilon_{\alpha_1 \beta_1}\epsilon_{\alpha_2 \beta_2}\epsilon_{\alpha\beta}\left(-X_{12}^{\alpha_1 \alpha_2}X_{23}^{\beta_2 \alpha}X_{13}^{\beta_1\beta}\right.\\
	&+\frac{1}{8}X_{11}^{\alpha_1 \beta_1}X_{22}^{\alpha_2 \beta_2}X_{33}^{\alpha\beta}-\frac{1}{4}X_{13}^{\alpha_1\alpha}X_{13}^{\beta_1\beta}X_{22}^{\alpha_2 \beta_2}\\
	&\left.-\frac{1}{4}X_{12}^{\alpha_1\alpha_2 }X_{12}^{\beta_1\beta_2}X_{33}^{\alpha\beta}-\frac{1}{4}X_{23}^{\alpha_2 \alpha}X_{23}^{\beta_2 \beta}X_{11}^{\alpha_1 \beta_1}\right)
\end{split}
\end{equation}
\end{minipage}

\vspace{\belowdisplayskip}
Observe that the components $X_{11}$, $X_{22}$ and $X_{33}$ of the meson $X$ correspond to $2 \times 2$ anti-symmetric matrix, i.e. they are singlets.  
The two $\USp{2}$ nodes have each $4$ fundamentals and are therefore confining \cite{Aharony:1997gp}. 
Here we choose to confine the $\USp{2}_1$ group. The other choice is completely equivalent
because of the $\SU{2}_A$ global symmetry that rotates the two anti-symmetric in the original description of the model (we will further comment on this equivalence below).
After confining the  $\USp{2}_1$  gauge group we are left with a $\USp{2}_2$ SQCD with  four fundamentals and a non-trivial superpotential. The quiver and the operator mapping  are reported below

\begin{minipage}[c]{0.3\linewidth}
\centering
\begin{tikzpicture}[auto, scale=1.5]
%%%%%%%%%%%%%% nodes %%%%%%%%%%
 	\node [circle,inner sep=0pt, minimum size=5mm] (N) at (0,0) {$2$};
    \node [rectangle, draw=red!50, fill=red!20, inner sep=0pt, minimum size=4mm] (F) at (2,0) {$2$};
    \node [cross out, draw=black, rotate=18, thick, scale=.7] (sing) at (0.6,0.265) {};
     %%%%%%%%%%%%% fields %%%%%%%%%%
    \draw ($(N.north east)!0.3!(N.south east)$) [bend left=30] to node {$X_{23}$} ($(F.north west)!0.3!(F.south west)$) [-,thick];
	\draw ($(N.south east)!0.3!(N.north east)$) [bend right=30, below] to node {$\Tilde{X}_{23}$} ($(F.south west)!0.3!(F.north west)$) [-,thick];
	\draw (F) to [out=-5, in=-60, looseness=8] (F) [-,thick];
    \draw (F) to [out=5, in=60, looseness=8] (F) [-, thick];
    \draw (N) to [out=175, in=120, looseness=8] (N) [-,thick];
    \draw (N) to [out=185, in=240, looseness=8] (N) [-, thick];
	
	\node [circle, draw=blue!50, fill=blue!20, inner sep=0pt, minimum size=5mm] (L) at (0,0) {$2$};
	\node [above left=0.3cm of N] (X22) {$X_{22}$};
	\node [below left=0.3cm of N] (X22t) {$\Tilde{X}_{22}$};
	\node [above right=0.3cm of F] (X33) {$X_{33}$};
	\node [below right=0.3cm of F] (X33t) {$\Tilde{X}_{33}$};
	\node [] (t) at (0.6,0) {$S_{1}$};
\end{tikzpicture}
\end{minipage}\hfill
\begin{minipage}[c]{0.5\linewidth}
\begin{equation}
\label{dualusp22ft}
\begin{split}
	\Tilde{X}_{23}^{\alpha_2\alpha}&=\epsilon_{\alpha_1 \beta_1}X_{12}^{\alpha_1\alpha_2}X_{13}^{\beta_1 \alpha}\\
	\Tilde{X}_{22}^{\alpha_2 \beta_2}&=\epsilon_{\alpha_1 \beta_1}X_{12}^{\alpha_1\alpha_2}X_{12}^{\beta_1\beta_2}\\
	\Tilde{X}_{33}^{\alpha\beta}&=\epsilon_{\alpha_1 \beta_1}X_{13}^{\alpha_1\alpha}X_{13}^{\beta_1\beta}\\
	S_{\scriptscriptstyle{1}}&=\epsilon_{\alpha_1 \beta_1}X_{11}^{\alpha_1 \beta_1}
\end{split}
\end{equation}
\end{minipage}\\
while the  superpotential is
\begin{equation}
\begin{split}
	W&=\epsilon_{\alpha_2\beta_2}\epsilon_{\alpha\beta}\left[Y_4\left(-\Tilde{X}_{23}^{\alpha_2\beta}X_{23}^{\beta_2\alpha}+\frac{1}{8}S_{1} X_{22}^{\alpha_2\beta_2}X_{33}^{\alpha\beta}-\frac{1}{4}\Tilde{X}_{33}^{\alpha\beta}X_{22}^{ab}\right.\right.\\
	&-\left.\left.\frac{1}{4}\Tilde{X}_{22}^{\alpha_2\beta_2}X_{33}^{\alpha\beta}-\frac{1}{4}S_{1}X_{23}^{\alpha_2\alpha}X_{23}^{\beta_2\beta}\right)+\frac{Y_2^{(1)}}{4}\left(\Tilde{X}_{22}^{\alpha_2\beta_2}\Tilde{X}_{33}^{\alpha\beta}-2\Tilde{X}_{23}^{\alpha_2\alpha}\Tilde{X}_{23}^{\beta_2\beta}\right)\right]
\end{split}
\end{equation}
The field $Y_2^{(1)}$ is the monopole of the $\USp{2}_1$ gauge group acting as a singlet in the confined phase.

We conclude the sequence by confining the $\USp{2}_2$ gauge group, that has indeed 
four fundamentals. This leads to the final, confined, theory where the new mesons 
are mapped to the fundamentals of  $\USp{2}_2$ as
\begin{equation}
\begin{array}{lll}
	V^{\alpha\beta}=\epsilon_{\alpha_2\beta_2}\Tilde{X}_{23}^{\alpha_2\beta}X_{23}^{\beta_2\alpha},\qquad &U^{\alpha\beta}=\epsilon_{\alpha_2\beta_2}X_{23}^{\alpha_2 \alpha}X_{23}^{\beta_2\beta},\qquad&T^{\alpha\beta}=\epsilon_{\alpha_2\beta_2}\Tilde{X}_{23}^{\alpha_2 \alpha}\Tilde{X}_{23}^{\beta_2\beta}
\end{array}
\end{equation}	
Furthermore there are two singlets of $\USp{2}_2$ that we redefine as $R_{1}=\epsilon_{\alpha_2\beta_2}X_{22}^{\alpha_2\beta_2}$ and $R_{2}=\epsilon_{\alpha_2\beta_2}\Tilde{X}_{22}^{\alpha_2\beta_2}$.
The superpotential of this final WZ model is 
\begin{equation}
\label{finalWZ}
\begin{split}
    W&=\epsilon_{\alpha\beta}\left[Y_{4}\left(-V^{\alpha\beta}+\frac{1}{8}S_{1}R_{1}X_{33}^{\alpha\beta}-\frac{1}{4}R_{1}\Tilde{X}_{33}^{\alpha\beta}-\frac{1}{4}R_{2}X_{33}^{\alpha\beta}-\frac{1}{4}S_{1}U^{\alpha\beta}\right)\right.\\
    &+\left.\frac{Y_2^{(1)}}{4}\left(R_2\Tilde{X}_{33}^{\alpha\beta}-2T^{\alpha\beta}\right)+\frac{Y_2^{(2)}}{2}\epsilon_{\ell m}\left(U^{\alpha \ell}T^{\beta m}-V^{\alpha\ell}V^{\beta m}\right)\right]
\end{split}
\end{equation}
where the field $Y_2^{(2)}$ is the monopole of the $\USp{2}_2$ gauge group acting as a singlet in the confined phase.
The expression (\ref{finalWZ})  needs some massage in order to simplify its interpretation.
For example some fields appear quadratically in the superpotential and they can be integrated out. By writing
\begin{equation}
    V_{\alpha\beta}=\sigma^{\mu}_{\alpha\beta}v_\mu,\qquad \sigma^\mu=(\mathbf{1},\sigma^i)   
\end{equation}
we see that the $v_3$ field is massive. The singlet field $\epsilon_{\alpha\beta}T^{\alpha\beta}$ also acquires a mass and it can be integrated out in the IR. By considering the various F-term conditions, we get the final superpotential
\begin{equation}
\label{WZWfin}
    W=Y_2^{(2)} \left[\frac{1}{2} R_2 U \Tilde{X}_{33}-\frac{1}{64}\left(R_1(S_1 X_{33}-2\Tilde{X}_{33})-2(R_2 X_{33}+S_1 U)\right)^2+\det V_{\alpha\beta}\right]
\end{equation}
We can identify the fields here with the ones in formula (\ref{OScharges2}) by first flipping the 
singlets $s_I$ in the original $\USp{4}$ gauge theory. This can be done by adding two fields, 
denoted as $r_I$ through the superpotential $\Delta W = r_I s_I$. These fields can be integrated out in the electric description and $F$-terms of $s_I$ leave us with the  identification $r_I \propto \Tr A_I^2$.
In the dual description the fields $r_I$ are crucial in order to reconstruct the correct field content  of the duality.

Looking at the global symmetry structure the explicit mapping is then
\begin{eqnarray}
    \label{mapping}
    Y_2^{(2)} &\leftrightarrow& \mathcal{T}_4 \nonumber \\
    X_{33} &\leftrightarrow& M \nonumber \\
    (S_1,R_1) &\leftrightarrow& (\phi_1,\phi_2) \nonumber \\
    (R_2,r_1,r_2) &\leftrightarrow& ( \phi_{12},\phi_{11},\phi_{22} ) \\
    V_{\alpha \beta} &\leftrightarrow& B_{\alpha \beta} \nonumber \\
    (\tilde{X}_{33},U)&\leftrightarrow& (B_{1},B_{2}) \nonumber 
\end{eqnarray}
Substituting this mapping into the superpotential (\ref{WZWfin}) we obtain
\begin{eqnarray}
W= \mathcal{T}_4  \Big(B_1 B_2 \phi _{1,2}-\big(\phi _2 \left(M \phi _1-B_1\right)-\left(M \phi _{1,2}+B_2 \phi _1\right)\big)^2 + \det B_{\alpha \beta}\Big)
\end{eqnarray}
where we absorbed the numerical coefficients into the fields.
Actually we could have reversed the order of the last  two confining dualities on $\USp{2}_1$ and  $\USp{2}_2$, arriving to a different results, with 
the role of $B_I$ and $\phi_I$ exchanged. However the two WZ models must be equivalent, and this equivalence corresponds to the following more symmetric formulation of the superpotential
\begin{eqnarray}
W&=& \mathcal{T}_4  \Big(B_1 B_2 \phi _{1,2}-\big(\phi _2 \left(M \phi _1-B_1\right)-\left(M \phi _{1,2}+B_2 \phi _1\right)\big){}^2\nonumber \\
&
-&\big(\phi _1 \left(M \phi _2-B_2\right)-\left(M \phi _{1,2}+B_1 \phi _2\right)\big){}^2+ \det B_{\alpha \beta}\Big)
\end{eqnarray}
The last step of the derivation of the superpotential of the WZ model consists of flipping $s_I$ in the electric model. This gives rise to the interactions involving the fields $\phi_{11}$
and $\phi_{22}$. By a symmetry argument we claim that the  interactions  allowed by the global symmetry are generated at quantum level and  that the flipped fields reconstruct the 
$\SU{2}_A$ adjoint field $\phi_{IJ}$.
In the next subsection we will confirm this expectation from the analysis of the partition function.

\subsection{3d partition function}
We complete our analysis by reproducing  the derivation of the duality  from supersymmetric localization on the squashed three sphere.
Such procedure gives rise to the identity between the partition function of $\USp{4}$ with 
with two anti-symmetric and two fundamentals and the partition function of the WZ model for the gauge singlets $B_{\alpha,\beta},B_I,\phi_I,\phi_{IJ},M$ and $\mathcal{T}_4$. The global symmetry enters in these identities in terms of real masses, that from the field theory side are associated to vevs of the reals scalars in the vector multiplets of the weakly gauged background flavor symmetries.

Before studying the deconfinement of two rank-$2$ anti-symmetric tensors from the three sphere partition function we  briefly review the necessary definitions. The partition function on the squashed three sphere $S_b^3$, obtained from localization in \cite{Hama:2011ea} (see also \cite{Kapustin:2009kz,Jafferis:2010un,Hama:2010av} for the round case) is a matrix integral over the reals scalar in the vector multiplet in the
Cartan of the gauge group. There is a classical term corresponding to the CS action (global and local) and the matter and the gauge multiplet contribute with their one loop determinant. These last can be associated to hyperbolic Gamma functions, formally defined as 
\begin{equation}
    \Gamma_h(z;\omega_1,\omega_2)=\prod_{n_1,n_2 \geq 0}^{\infty}
\frac{(n_1+1)\omega_1+(n_2+1)\omega_2- z}
{n_1 \omega_1+n_2 \omega_2+z}
\end{equation}
The argument of such Gamma functions is physically interpreted as a holomorphic combination between the real masses for the gauge and the global symmetries and the R-charges (or mass dimensions).
The purely imaginary parameters $\omega_1=i b$ and $\omega_2=i/b$ are related to the squashing parameter of the three sphere $S_b^3$.

Here we will only focus on the case of symplectic gauge group. Let us consider the partition function of an  $\USp{2 N_c}$ gauge theory with $2 N_f$ fundamentals. It  is given by
\begin{eqnarray}
\label{uspele}
	Z_{USp(2N_c),N_f}(\mu) &=& \frac{1}{2^n n!  \sqrt{(-\omega_1 \omega_2)^{n}}} \int  \prod_{i=1}^{N_c}  \dd{z_i}  \frac{\prod_{a=1}^{2N_f} \Gamma_h(\pm z_i + \mu_a)}{\Gamma_h(\pm 2 z_i)}\prod_{i<j} \frac{1}{\Gamma_h(\pm z_i \pm z_j  )} \nonumber \\
\end{eqnarray}
In our analysis we will use an identity involving this partition function and its dual Aharony phase \cite{Aharony:1997gp}.
The identity is (see Theorem 5.5.9 of \cite{vanDeBult})
\begin{eqnarray}
\label{ahausp}
Z_{USp(2N_c),N_f}(\mu) 
&=&
\Gamma_h\left(2\omega(N_f-N_c))-\sum_{a=1}^{2N_f} \mu_a\right) \nonumber \\
&\times &
\prod_{a<b} \Gamma_h(\mu_a +\mu_b)
Z_{USp(2(N_f-N_c-1)),N_f}(\omega-\mu) 
\end{eqnarray}
with $2 \omega \equiv \omega_1 +\omega_2$.
Observe that the identity (\ref{ahausp}) remains  valid  for $N_f=N_c+1$, that corresponds to the confining case of Aharony duality \cite{Aharony:1997gp}, where only the meson $M$ and the minimal $\USp{2N_c}$ monopole $Y$ survive in the WZ model and they interact through the superpotential  $W=Y \Pf{M}$.

We start considering the original model, adding also the flippers $s_I$ arising from the superpotential (\ref{Wflip}).
The partition function is
\begin{eqnarray}
\label{startingZ}
	Z &=& \frac{\prod_{A=1,2}\Gamma_h(2\omega-2n_{A}) }{8 \sqrt{-\omega_1 \omega_2}^2} \int  \prod_{i=1,2}  \dd{z_i}
	\frac{\prod_{a=1,2}\Gamma_h(\pm z_i +m_a)}{\Gamma_h(\pm 2 z_i )} 
	%\nonumber \\ &\times& 
	\frac{\prod_{A=1,2}\Gamma_h(\pm z_1 \pm z_2 +n_{A})}{\Gamma_h(\pm z_1 \pm z_2 )} \nonumber \\ 
\end{eqnarray}
In this formula $m_{1,2}$ are the real masses of the two fundamental fields and $n_{1,2}$ are the real masses of the two anti-symmetric fields.
We can also use a different basis
\begin{equation}
\label{nonab}
m_1 = \rho+\sigma,\quad
m_2 = \rho-\sigma,\quad
n_1 = \mu+\nu,\quad 
n_2 = \mu-\nu
\end{equation}
giving an explicit parameterization it terms of the Cartan of the $\U{2}^2$ flavor symmetry.
Indeed in this way $\sigma$ and $\nu$ parameterize the Cartan of $\SU{2}_a$ and $\SU{2}_A$ respectively.

We then proceed by deconfining the two rank-$2$ anti-symmetric tensors. This step produces two 
$\USp{2}$ gauge nodes, two bifundamentals, each connecting one of these $\USp{2}$ gauge groups to the original $\USp{4}$. 
The partition function of the model becomes
\begin{eqnarray}
\label{decZ}
Z &=& \frac{1}{32 \sqrt{(-\omega_1 \omega_2)^{4}}} \int  
\frac{\dd{z_1}\dd{z_2} \dd{w_1} \dd{w_2}}{
\Gamma_h(\pm 2z_1 )
\Gamma_h(\pm 2z_2 )
\Gamma_h(\pm 2w_1 )
\Gamma_h(\pm 2w_2 )
}
\nonumber \\
&\times&
\prod_{i=1,2} \left(
\prod_{a=1,2}\Gamma_h(\pm z_i +m_a)
\cdot
\prod_{A=1,2}\Gamma_h(\pm z_i \pm w_A +n_{A}/2)\right)
\end{eqnarray}
As a check we can see that  (\ref{startingZ}) is obtained by applying (\ref{ahausp}) to the two  $\USp{2}$ gauge groups in (\ref{decZ}).
The partition function  (\ref{decZ})  corresponds to the one for the model in represented in (\ref{decft}).

The next step consists of Aharony duality on $\USp{4}$. At the level of the partition function it corresponds to use the identity (\ref{ahausp}) on the gauge theory identified by the variables $z_{1,2}$.
The partition function becomes
\begin{eqnarray}
\label{dualusp4}
Z &=& \frac{1}{4  \sqrt{(-\omega_1 \omega_2)^{2}}}
\Gamma_h(2\omega - m_1-m_2-n_1-n_2)\Gamma_h(m_1+m_2)\prod_{A=1,2}\Gamma_h(n_A)
\nonumber \\
&\times&
\int \dd{w_1} \dd{w_2} 
\frac{\prod_{A=1,2}
\Gamma_h(m_a\pm w_A+n_A/2) \cdot \Gamma_h(\pm w_1\pm w_2+(n_1+n_2)/2)
}{
\Gamma_h(\pm 2w_1 )
\Gamma_h(\pm 2w_2 )
}
\nonumber \\
\end{eqnarray}
The partition function  (\ref{dualusp4})  corresponds to the one for the model in represented in (\ref{dualusp4ft}).

The next step consists of a confining limit of Aharony duality on one of the $\USp{2}$ factor. Choosing one of the two  $\USp{2}$ nodes has the effect of making the $\SU{2}_A \times \SU{2}_a$ global symmetry  not  manifest in the integrand of the partition function.
Following the discussion on the field theory side here we choose to dualize the  $\USp{2}_1$ gauge group, such that the partition function becomes 
\begin{eqnarray}
    \label{dualusp22}
    Z &=& \frac{1}{2} 
    \Gamma_h(2\omega - m_1-m_2-n_1-n_2)
    \Gamma_h(m_1+m_2)
    \prod_{A=1,2}\Gamma_h(n_A)
    \nonumber \\
    &\times&
    \Gamma_h(m_1+m_2+n_1)
    \Gamma_h(n_1+n_1)
    \Gamma_h(2\omega-2n_1-n_2-m_1-m_2)
    \nonumber \\
    &\times&
    \int \dd{w_2} 
    \frac{
    \prod_{a=1,2}
    \Gamma_h(m_a\pm w_2+n_2/2+n_1)\cdot 
    \prod_{A=1,2}
    \Gamma_h(m_a\pm w_2+n_2/2)}{\Gamma_h(\pm 2w_2 )}
    \nonumber \\
\end{eqnarray}
The partition function  (\ref{dualusp22})  corresponds to the one for the model in represented in (\ref{dualusp22ft}).

The last step of the procedure requires a confining limit of Aharony  duality on the leftover  
$\USp{2}_2$ gauge group. This gives the final partition function
\begin{eqnarray}
\label{damassaggiare}
Z &=& 
\Gamma_h(2\omega - m_1-m_2-n_1-n_2)
\Gamma_h(m_1+m_2)
\prod_{A=1,2}\Gamma_h(n_A)
\nonumber \\
&\times&
\Gamma_h(m_1+m_2+n_1)
\Gamma_h(n_1+n_2)
\Gamma_h(2\omega-2n_1-n_2-m_1-m_2)
\nonumber \\
&\times&
\Gamma_h(m_1+ m_2+2n_1+n_2)
\Gamma_h(m_1+ m_2+n_1)
\nonumber \\
&\times&
\Gamma_h(2\omega-2m_1-2m_2-2n_1-2n_2)
\prod_{a,b=1,2}
\Gamma_h(m_a+m_b+n_1+n_2)
\end{eqnarray}

This expression still needs some massage. First we can integrate out the massive fields, as done on the field theory approach. Here this integration corresponds to take advantage of the formula 
$\Gamma_h(2 \omega-x)\Gamma_h(x)=1$.
After this step we can also write down (\ref{damassaggiare}) in a manifestly  $\SU{2}_A \times \SU{2}_a$ invariant form. We arrive to the expression
\begin{eqnarray}
    Z &=& 
    \Gamma_h(m_1+m_2)\Gamma_h(n_1+n_2) \prod_{A=1,2}(\Gamma_h(n_A) \, \Gamma_h(m_1+m_2+n_A))
    \nonumber \\
    &\times&
    \Gamma_h(2\omega-2m_1-2m_2-2n_1-2n_2)
    \prod_{a \leq b}
    \Gamma_h(m_a+m_b+n_1+n_2)
    \end{eqnarray} 
    or using (\ref{nonab})
    \begin{eqnarray}
    \label{final}
    Z &=& 
    \Gamma_h(2 \rho)
    \Gamma_h(2\mu) 
    \Gamma_h(\mu \pm \nu) 
    \Gamma_h(2\rho + \mu \pm \nu)
    \nonumber \\
    &\times&
    \Gamma_h(2\omega-4 \mu - 4 \rho)
    \Gamma_h(2 \rho \pm 2 \sigma+2\mu,2 \rho+2\mu)
\end{eqnarray}
This is the final expression that matches with (\ref{startingZ}).
We can also flip the singlets $s_I$ in the electric side, and on the magnetic side two new singlets 
appear with their contribution to the  $\Gamma_h(2 n_A) = \Gamma_h(2 \mu \pm 2 \nu)$ to the partition function.
In this way we can see that all the fields $B_I$, $\phi_I$, $\phi_{IJ}$, $B_{\alpha,\beta}$,
$M$ and the monopole $\mathcal{T}_4$ appear in the partition function with the expected real masses. 
Explicitly we can associate these Gamma functions to the singlets of the confined phase using the mapping
\begin{eqnarray}
    &B_{\alpha \beta} \leftrightarrow    \Gamma_h(2 \rho \pm 2 \sigma+2\mu,2 \rho+2\mu) \quad &\phi_{IJ}\leftrightarrow  \Gamma_h(2 \mu \pm 2 \nu,2\mu) \nonumber\\
    & \phi_{I} \leftrightarrow  \Gamma_h(\mu \pm \nu) \quad &B_{I} \leftrightarrow  \Gamma_h(2\rho + \mu \pm \nu) \\
    &M \leftrightarrow  \Gamma_h(2 \rho) \quad &\mathcal{T}_4 \leftrightarrow \Gamma_h(2\omega-4 \mu - 4 \rho)\nonumber 
\end{eqnarray}
Indeed the arguments of hyperbolic Gamma  functions correspond to the real masses that can be read from the charges in formula (\ref{OScharges2}).

\section{Conclusions}
In this paper we have derived, using field theory arguments, two confining dualities 
that have been proposed in the literature from supersymmetric localization. Here the dualities have been derived by combining the technique of rank-$2$  tensor deconfinement of \cite{Berkooz:1995km} together with the sequential application of ordinary dualities and/or confining dualities.

There are many interesting directions that would be worth to explore.

For example the Higgsing triggered, in the 4d model, by the quantum correction imposed on the moduli space, corresponds, on the superconformal index, to the pole pinching \cite{Gaiotto:2012xa} vastly used in \cite{Bajeot:2022kwt} for the derivation of 4d confining dualities  in presence of a superpotential. The confining theories obtained in this way were not discussed in 
\cite{Csaki:1996sm}, because of the absence of a superpotential. On the other hand 
confining gauge theories of this type have been discussed in \cite{Klein:1998uc}, at least for  the limiting cases of ordinary dualities with rank-$2$  tensor matter fields.
Therefore we expect many more 4d confining gauge theories with a simple gauge group  not  discovered yet.

Another interesting question regards the 3d duality for $\USp{4}$  with two rank-$2$ anti-symmetric tensors
and two fundamentals. As discussed in \cite{Okazaki:2023hiv} the field content in this case corresponds to the one of the dualities studied in \cite{Amariti:2022iaz,Amariti:2022lbw} with D-type superpotential. Nevertheless as observed in \cite{Okazaki:2023hiv} there are differences in the operator mapping and in the charge spectrum.
Furthermore the $\USp{4}$ duality discussed here appears sporadic and its generalization to $\USp{2N_c}$ does not seem straightforward. For example we did not find any confining duality by increasing the rank of the gauge group and keeping fixed the field content (i.e. keeping  two rank-$2$ antisymmetric tensors and possibly increasing the number of fundamentals).
It is nevertheless possible that further fields and interactions should be considered in order to have an $\USp{2N_c}$ confining theory with two rank-$2$ anti-symmetric tensors.

A last, related, question regards the existence of 4d confining dualities with two rank-$2$ tensors. Beyond the case of $\USp{2N_c}$  with two rank-$2$ anti-symmetric tensors, one can imagine also cases with unitary or orthogonal gauge groups or cases with more general rank-$2$ tensor matter fields.

%%%%%%%%%%%
\section*{Acknowledgments}
%%%%%%%%%%%%%%%%%%
%
%
We are grateful to Sara Pasquetti and Simone Rota for discussions.
D.M thanks the Perimeter Institute for Theoretical Physics for the hospitality and the organizers of the ”Strings 2023” conference, during which this work has been completed. The work of A.A., D.M. has been supported in part by the Italian Ministero dell'Istruzione, Universit\`a e Ricerca (MIUR), in part by Istituto Nazionale di Fisica Nucleare (INFN) through the “Gauge Theories, Strings, Supergravity” (GSS) research project and in part by MIUR-PRIN contract 2017CC72MK-003. 

\appendix

\section{Remarks on the 4d index}
\label{appendixA}
In this appendix we collect the mathematical identities that have been useful in our analysis 
of the  SWV duality. Such identities correspond to the matching of the electric and the magnetic index for the 4d duality of \cite{Seiberg:1994pq} and of \cite{Intriligator:1995ne}. 
 The identities hold also in the s-confining case,  when the dual theory become a WZ model.
 
 Skipping the definitions and the conventions that we use for the index (that correspond to the ones of \cite{Dolan:2008qi,Spiridonov:2009za}) where the relevant quantities are
the elliptic gamma functions and the  Pochammer symbols
\begin{equation}
    \Gamma_e(z;p,q)=\prod_{\ell,m=1}^\infty\frac{1-z^{-1}p^{\ell+1}q^{m+1}}{1-z p^\ell q^m},\qquad     (x,p)_\infty=\prod_{\ell=0}^\infty(1-xp^\ell) \\
\end{equation}
here we provide the integral identities matching the indices across duality.
 
 In the case of $\SU{N_c}$ SQCD with $N_f$ flavors Seiberg duality corresponds on the supersymmetric index to the integral identity between
 \begin{equation}
    \begin{aligned}
    I_E= & \frac{(p ; p)_{\infty}^{N_c-1}(q ; q)_{\infty}^{N_c-1}}{N_c !} \\
    & \quad \times \int_{\mathbb{T}^{N_c-1}} \frac{\prod_{i=1}^{N_f} \prod_{j=1}^{N_c} \Gamma_e\left(s_i z_j, t_i^{-1} z_j^{-1} ; p, q\right)}{\prod_{1 \leq i<j \leq N_c} \Gamma_e\left(z_i z_j^{-1}, z_i^{-1} z_j ; p, q\right)} \prod_{j=1}^{N_c-1} \frac{\dd{z_j}}{2 \pi \mathrm{i} z_j}
    \end{aligned}
    \label{SDele}
\end{equation}
and 
\begin{equation}
    \begin{aligned}
    I_M= & \frac{(p ; p)_{\infty}^{\widetilde{N}_c-1}(q ; q)_{\infty}^{\widetilde{N}_c-1}}{\widetilde{N}_c !} \prod_{1 \leq i, j \leq N_f} \Gamma_e\left(s_i t_j^{-1} ; p, q\right) \\
    & \quad \times \int_{\mathbb{T}^{\widetilde{N}_c-1}} \frac{\prod_{i=1}^{N_f} \prod_{j=1}^{\widetilde{N}_c} \Gamma_e\left(S^{1 / \widetilde{N}_c} s_i^{-1} z_j, T^{-1 / \widetilde{N}_c} t_i z_j^{-1} ; p, q\right)}{\prod_{1 \leq i<j \leq \widetilde{N}_c} \Gamma_e\left(z_i z_j^{-1}, z_i^{-1} z_j ; p, q\right)} \prod_{j=1}^{\widetilde{N}_c-1} \frac{\dd{z_j}}{2 \pi \mathrm{i} z_j}
    \end{aligned}
    \label{SDmag}
\end{equation}
where $S=\prod_{i=1}^{N_f}s_i$, $T=\prod_{i=1}^{N_f}t_i$ and $\widetilde{N}_c=N_f-N_c$. The equality holds with the following constraint on the fugacities $ST^{-1}=(pq)^{N_f-N_c}$.

 Observe that the relation between (\ref{SDele}) and (\ref{SDmag}) holds also for $N_f=N_c+1$ where the integral on the RHS vanishes. This provides the relation for the s-confining limit of Seiberg duality.
  
In the case of $\USp{2N_c}$ SQCD with $2N_f$ fundamentals  the index is given by
\begin{eqnarray}
    \label{usp}
    I_{\USp{2N_c},2N_f}(t) &=&\frac{(p,p)^{N_c}_{\infty} (q,q)^{N_c}_{\infty}}{2^{N_c} N_c!} \int_{\mathbb{T}^{N_c}} \prod_{1 \leq u < v \leq N_c} \frac{1}{\Gamma_e(x_u^{\pm 1} x_v^{\pm 1}; p,q)} \nonumber \\
    &\times &
    \prod_{u=1}^{N_c} 
    \frac{\prod_{i=1}^{2N_f} 
    \Gamma_e(t_i x_u^{\pm 1}; p,q)}{\Gamma_e(x_u^{\pm 2}; p,q)} \prod_{u=1}^{N_c} \frac{\dd{x_u}}{2 \pi i x_u}
\end{eqnarray}
and Intriligator-Pouliot duality corresponds to the integral identity
\begin{equation}
\label{uspid}
 I_{\USp{2N_c},2N_f}(t) 
 =
 \prod_{i<j} \Gamma_e(t_i t_j) \,
  I_{\USp{2(N_f-N_c-4)},2N_f}(pq/t) 
\end{equation}
where the fugacities are constrained by $\prod_{i=1}^{2N_f} t_i = (pq)^{N_f-N_c-1}$
Observe that the relation (\ref{uspid}) holds also for $N_f=N_c+2$ where the integral on the RHS vanishes. This provides the relation for the s-confining limit of Intriligator-Pouliot duality.

 Another identity that played a crucial role in our derivation was obtained in \cite{Spiridonov:2014cxa} for the case of $\USp{2N_c}$ SQCD with $2N_c+2$ fundamentals.  
 The identity in this case is 
 \begin{equation}
 \label{A6}
    \begin{split}
       & I_{\USp{2N_c},2N_c+2}\left(e^{2\pi i \phi} \right)  
         = \frac{1}{(p;p)_{\infty}^{N_c} (q,q)^{N_c}_{\infty} } \\ \times & \sum_{(\Phi_1 \bigcup \Phi_2)/ S_2^k}  \prod_{1 \leq i < j \leq N_c+1} \Gamma(e^{2 \pi i (\pm \tilde {\phi}_i \pm \tilde {\phi}_j )};p,q) \sum_{S_{N_c+1} (\Phi_2)} \prod_{i=1}^{N_c} \delta (\Tilde{\phi}_i + \Tilde{\phi}_{N_c+1+i} ),
     \end{split}
\end{equation}
with 
 $z_i = e^{ 2 \pi i \phi_i}$,  $\sum_{i=1}^{2N_c+2} \phi_i = 0$, $\Phi_1 = (\tilde{\phi}_1, ... , \tilde{\phi}_{N_c}, \tilde{\phi}_{k+1} = \phi_{N_c+1})$ and $\Phi_2 = (\tilde{\phi}_{N_c+2}, ... , \tilde{\phi}_{2N_c+2})$.
This relation reflects the statement that the theory confines with a quantum corrected moduli space.

\bibliographystyle{JHEP}
\bibliography{ref.bib}
\end{document}